%% file: 2021_conference.tex
\documentclass{article} % For LaTeX2e
\usepackage{2021_conference,times}
\usepackage{times}
\usepackage{epsfig}
\usepackage{graphicx}
\usepackage{amsmath}
\usepackage{amssymb}
\usepackage{wrapfig}
\usepackage{graphicx}
\usepackage{color}

\input{math_commands.tex}

\usepackage{hyperref}
\usepackage{url}

\title{The Benefit of Distraction: Denoising Remote Vitals Measurements using Inverse Attention}

\author{Ewa M. Nowara$^{1}$ Daniel McDuff$^{2}$ and Ashok Veeraraghavan$^{1}$ \\
Rice University$^{1}$, Microsoft Research AI$^{2}$ \\
}

\newcommand{\Section}[1]{\vspace*{-11pt}\section{#1}\vspace*{-11pt}}
\newcommand{\SubSection}[1]{\vspace*{-9pt}\subsection{#1}\vspace*{-9pt}}

\iclrfinalcopy % Uncomment for camera-ready version, but NOT for submission.
\begin{document}

\maketitle

\vspace*{-25pt}

\begin{abstract}
Attention is a powerful concept in computer vision. End-to-end networks that learn to focus selectively on regions of an image or video often perform strongly. However, other image regions, while not necessarily containing the signal of interest, may contain useful context. We present an approach that exploits the idea that statistics of noise may be shared between the regions that contain the signal of interest and those that do not. Our technique uses the inverse of an attention mask to generate a noise estimate that is then used to denoise temporal observations. We apply this to the task of camera-based physiological measurement. A convolutional attention network is used to learn which regions of a video contain the physiological signal and generate a preliminary estimate. A noise estimate is obtained by using the pixel intensities in the inverse regions of the learned attention mask, this in turn is used to refine the estimate of the physiological signal. We perform experiments on two large benchmark datasets and show that this approach produces state-of-the-art results, increasing the signal-to-noise ratio by up to 5.8 dB, reducing heart rate and breathing rate estimation error by as much as 30\%, recovering subtle pulse waveform dynamics, and generalizing from RGB to NIR videos without retraining.
\end{abstract}

\Section{Introduction}
Attention mechanisms have been successfully applied in many areas of machine learning and computer vision~\citep{mnih2014recurrent,vaswani2017attention}, including object detection~\citep{oliva2003top}, activity recognition~\citep{sharma2015action}, language tasks~\citep{anderson2018bottom,you2016image}, machine translation~\citep{bahdanau2014neural}, and camera-based physiological measurement~\citep{chen2018deepphys}. An additional benefit of attention mechanisms is that they are interpretable and show which regions of an image were used to generate a particular output. In this paper, we focus on a counter-intuitive question -- is there important information contained within the regions that are typically ignored by the attention models? And, can we exploit information in this region to improve the quality of estimation for the underlying signals of interest?

We focus on the specific temporal prediction problem of camera-based physiological measurement as an exemplar application for our approach. The SARS-CoV-2 (COVID-19) pandemic has rapidly changed the face of healthcare, emphasizing the need for better technology to remotely provide care to patients.
COVID-19 is linked to serious heart and respiration related symptoms~\citep{xu2020pathological,zheng2020covid,puntmann2020outcomes}.
Even after the COVID-19 crisis, many doctor appointments could be carried out online with telemedicine technology, increasing the flexibility for appointments. Recent research in computer vision has led to the development of non-contact physiological measurement techniques that leverage cameras and computer vision algorithms~\citep{takano2007heart, verkruysse2008remote, poh2010non, CHROMdeHaan, wang2017algorithmic, chen2018deepphys}. Camera-based vital signs could also enable driver monitoring~\citep{nowara2018sparseppg}, face anti-spoofing~\citep{liu2020temporal, nowara2017ppgsecure}, or long-term human-computer-interaction (HCI) studies~\citep{mcduff2016cogcam} where wearing contact devices for extended periods may be infeasible. Convolutional networks currently provide state-of-the-art performance on heart rate (HR) and breathing rate (BR) measurement from video~\citep{chen2018deepphys,yu2019remote,liu2020multi}.

While the convolutional neural networks may be able to accurately learn what features in the image are important for finding the physiological signals, they may not be able to learn a good model of the noise that corrupts the signals. The noise present in the video, which is considered to be ``everything else than the signal of interest", may be caused by many diverse factors and could vary greatly across videos and datasets. Possible sources of noise include changes in head motion~\citep{estepp2014recovering}, facial expressions~\citep{zhang2016multimodal}, speech, ambient light variations~\citep{nowara2018sparseppg}, and video compression artifacts~\citep{yu2019remote,nowara2019combating}. The wide variety of possible noise sources makes it challenging for any model to explicitly capture a good noise representation and to remove that noise from the signals of interest.

The key observation we make is that regions ignored by an attention mechanism in a neural model likely contain information about sources of noise that are also present in the regions used by the attention mechanism to compute the physiological signals. Using the ``distraction" regions that were ignored by the attention masks offers a way to estimate the noise for each video without making any assumptions about the nature of the noise. The only assumption is that most regions not used by the attention masks do not contain the signals of interest and consequently contain noise.

\begin{figure*}[t!]
  \includegraphics[width=1\textwidth]{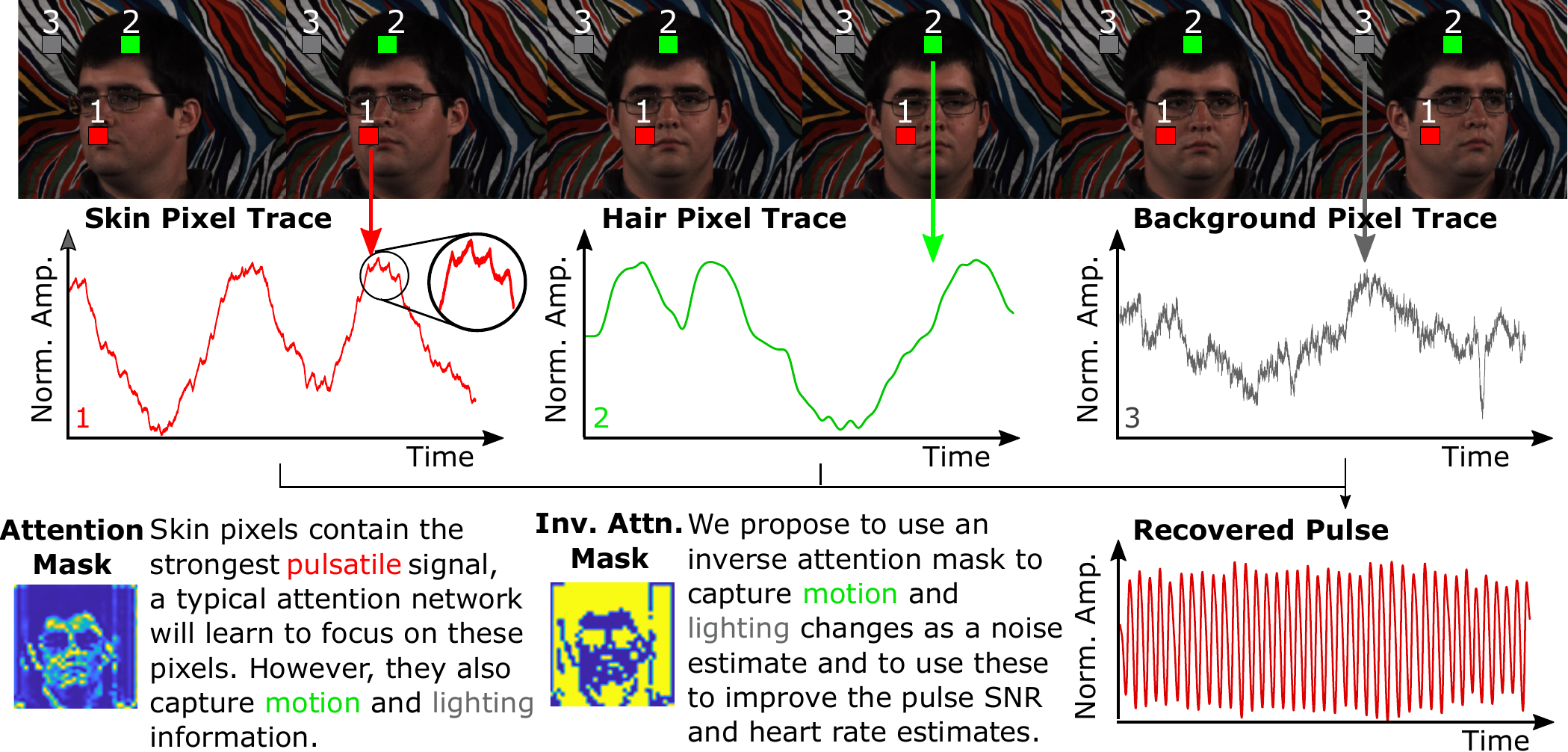}
  \caption{We propose an approach, using the regions ignored by the attention mechanism, that produces more accurate physiological waveforms, even in severely challenging scenarios.} %Attention mechanisms focusing on regions of interest have been shown to be useful for extracting physiological signals from videos. In severely challenging scenarios, where motion or illumination induce structured noise to the video data, these methods alone are insufficient. We propose an approach, using the concept of inverse attention, that produces more accurate pulse waveforms and heart rate estimates.}
  \label{fig:overview}
  \vspace{-0.5cm}
\end{figure*}

\begin{figure*}[t!]
  \includegraphics[width=1\textwidth]{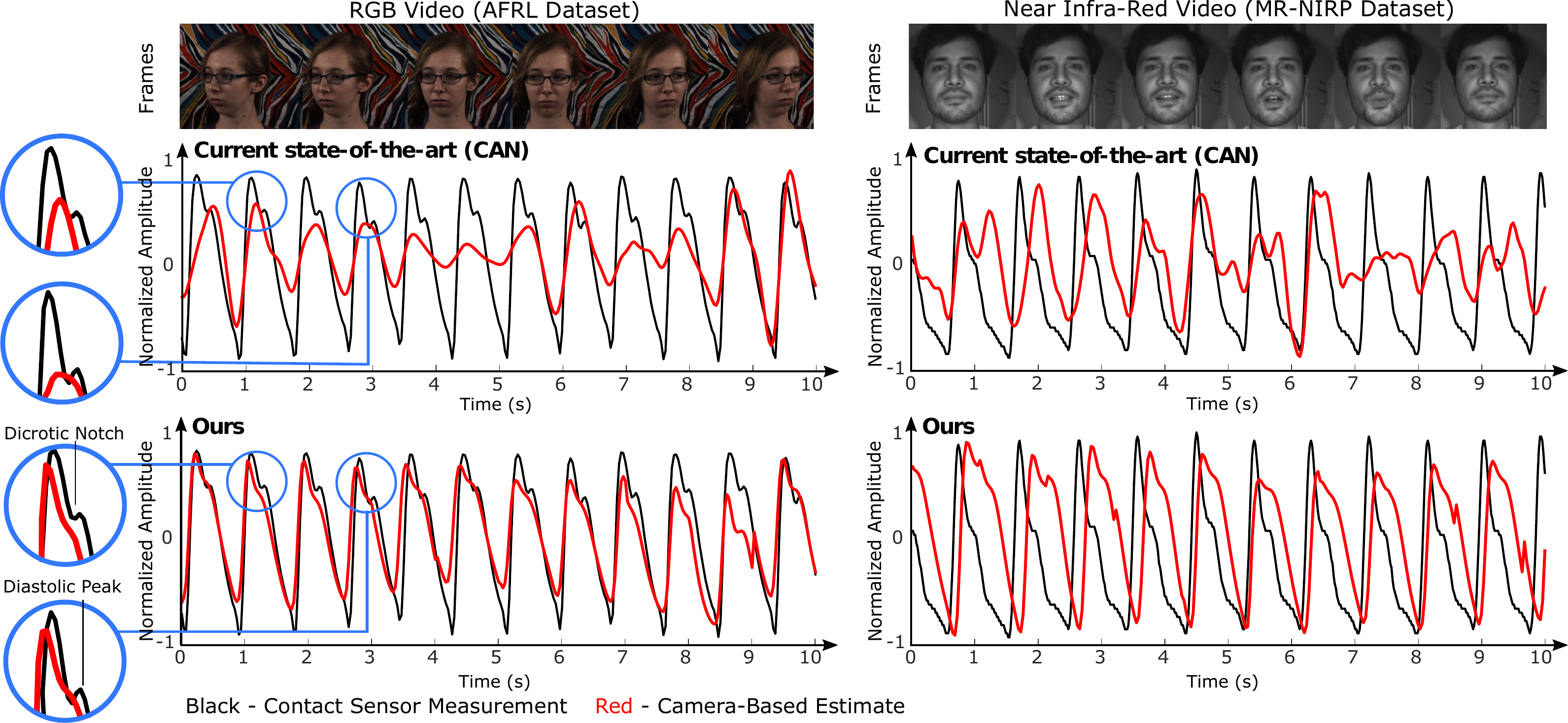}
  \caption{Pulse signals output by a state-of-the-art network and our denoising method. Our method produces cleaner signals, free from motion artifacts (present in the benchmark method), and better matching the ground truth subtle dynamics and shape. Notice the zoomed-in portions with easily identifiable dicrotic notch and diastolic peaks in our outputs.} 
  \label{fig:waveforms}
  \vspace{-0.5cm}
\end{figure*}

We demonstrate that we can use the intensity variations from regions outside of the attention mask as a noise estimate and learn a denoising mapping to remove noise from the recovered signals. See Fig.~\ref{fig:overview} for an overview of our denoising approach. We show that our approach outperforms state-of-the-art methods on several datasets across a range of HR and BR error measures. Our denoising approach also generalizes well to new data, even data recorded with different imaging modalities, such as near-infrared (NIR), without any additional training. Our proposed approach is also able to recover very subtle waveform dynamics, such as the clearly visible dicrotic notch, shown in Fig.~\ref{fig:waveforms}, which is challenging for video-based methods. Obtaining clean and more accurate waveforms is useful for determining important health metrics, such as blood pressure~\citep{elgendi2019use}, which is infeasible with current methods. The idea of using the inverse attention regions is likely very useful in a wide variety of vision tasks, ranging from activity recognition to deblurring. However, in this work, we focus on physiological measurement due to the clinical importance.

The core contributions of this paper are to: (1) propose the use of inverse attention masks for generating noise estimates, (2) present a novel method for denoising non-contact physiological measurement using this approach, (3) evaluate our method on three datasets showing state-of-the-art performance on pulse and respiration measurement, (4) demonstrate that our approach generalizes to NIR data without further training. Supplementary material including code, models, video examples and additional experimental results are provided with this submission.\footnote{\url{https://github.com/AnonymousCodeSubmission/Benefit_of_Distraction}}

\Section{Related Work}
\textbf{Attention Mechanisms.} Attention mechanisms provide a way for a model to learn which parts of an image or video ``are relevant for the task at hand and attach a higher importance to them''~\citep{sharma2015action}. During training the attention weights are learned reflecting the importance of the embedding features. Recently, transformer models, based solely on attention mechanisms, have become popular~\citep{vaswani2017attention}. In convolutional neural networks (CNNs) these attention mechanisms typically form a spatial mask. These masks can help practitioners understand the decision-making process of a network~\citep{fukui2019attention} and in certain cases the ``fixations'' of attention generated by computer models and by human observers were very similar~\citep{oliva2003top}. Attention mechanisms can be used to connect layers; for example, one which focuses on temporal information (e.g., trained on flows) and another which focuses on spatial information (e.g., trained on RGB frames). Prior work has found that these crosslink layers guide the spatial-stream to pay more attention to the human foreground areas and can be less affected by background clutter~\citep{tran2017two}. In physiological measurement, two-layer networks have been found to be effective as both color and motion information are valuable for extracting the subtle physiological signal in the presence of noise~\citep{chen2018deepphys}. While attention mechanisms often work well, they are a simple representation of which regions are important. However, pixels outside these regions may provide useful context or a strong prior about the noise present.

\textbf{Physiological Imaging.} Volumetric changes in blood over time lead to subtle changes in light reflected from the skin and subtle motion variations which can be measured with a camera~\citep{takano2007heart,verkruysse2008remote}. The physiological signal obtained from a video can be used to recover several metrics and vital signs, including heart rate~\citep{poh2010non}, heart rate variability~\citep{poh2010advancements}, breathing rate~\citep{poh2010advancements}, blood oxygenation~\citep{tarassenko2014non} and pulse transit time~\citep{shao2014noncontact}. NIR~\citep{nowara2018sparseppg, chen2018estimating} and thermal cameras have also been successfully used for measuring physiological signals in the dark~\citep{garbey2007contact, pavlidis2016dissecting}. Unfortunately, the signals of interest in video-based physiological measurement are often very subtle and can be easily corrupted by noise due to body motions and ambient lighting changes. Early work in physiological imaging used properties of the physiological signal, e.g., the periodic nature~\citep{poh2010non} and hemoglobin absorption spectra~\citep{CHROMdeHaan,wang2017algorithmic} to recover the underlying physiological signal via de-mixing methods. Others have used physical skin models to learn a mapping from color changes~\citep{mcduff2018fast}. %These methods all involve multi-stage methods including face detection and segmentation, spatial averaging and color channel de-mixing (signal source separation)~\citep{poh2010non,CHROMdeHaan,wang2017algorithmic,nowara2018sparseppg,tulyakov2016self}. It is well established that end-to-end deep neural models can outperform traditional multi-stage methods that require hand-crafted features. Deep-learning has been similarly successful in extracting the BVP from video~\citep{chen2018deepphys,mcduff2018deep,vspetlik2018visual,zhan2019analysis}. 
Recently, several groups have demonstrated that deep learning models free from heuristic assumptions about the signal structure can perform better, especially in presence of large motion and noise~\citep{chen2018deepphys,zhan2019analysis,vspetlik2018visual,mcduff2018deep,niu2018synrhythm}. We show that the performance of a state-of-the-art model is significantly improved by using the distraction regions as a noise estimate. 
%Chen et al.~\citep{chen2018deepphys} used CANs to produce BVP and HR estimates, and masks of regions in the image that were used for computing the signals. This provides interpretability of which pixels were clean or noisy in a given video. The output attention masks learn a form of skin segmentation combined with blood perfusion. %Moreover, deep learning methods are capable of learning complex relationships that are difficult to model with heuristic methods, for example recovering pulse estimates from heavily compressed videos~\citep{yu2019remote,nowara2019combating}. 

\Section{Benefiting from Distraction}
\begin{figure*}[t!]
  \includegraphics[width=1\textwidth]{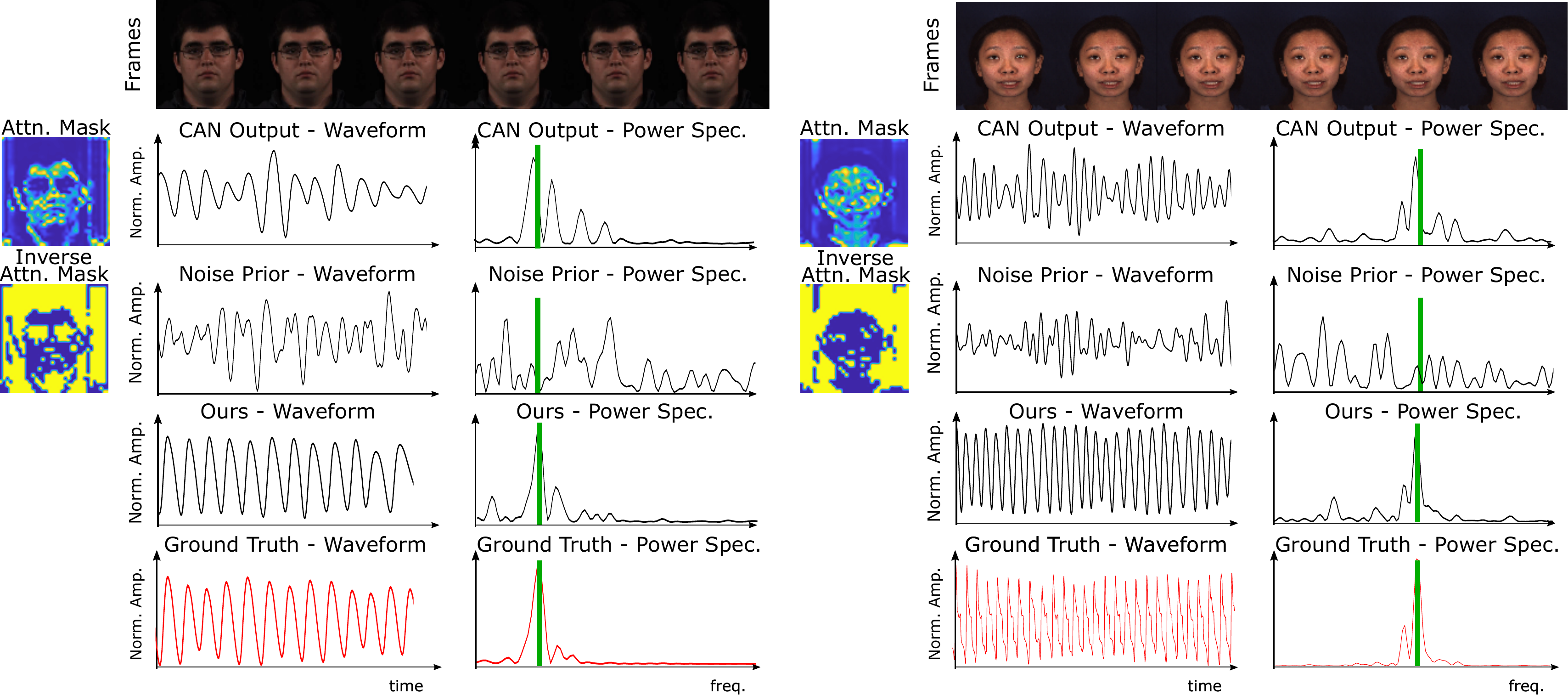}
  \caption{Examples of attention masks, inverse attention masks (yellow=higher weight) and the corresponding original physiological estimates, the noise estimates (shown for the green channel only for visual clarity), and denoised physiological signals from AFRL (left) and MMSE-HR (right) datasets. Green lines indicate the ground truth HR.} %Notice how our method produces cleaner waveforms and more accurate peak estimates.} 

%   \dan{Ewa, you are not showing the RGB noise estimates here that will confuse people - which channel did it come from?}
  \label{fig:masks}
\end{figure*}
% We hypothesize that regions ignored or given low weight by an attention mechanism may be helpful in generating noise estimates and thus be leveraged to enhance the quality of the signal of interest. In this work, we consider the application of measuring physiological signals from video to test this hypothesis and demonstrate the benefits. 

\begin{figure*}[t!]
  \includegraphics[width=1\textwidth]{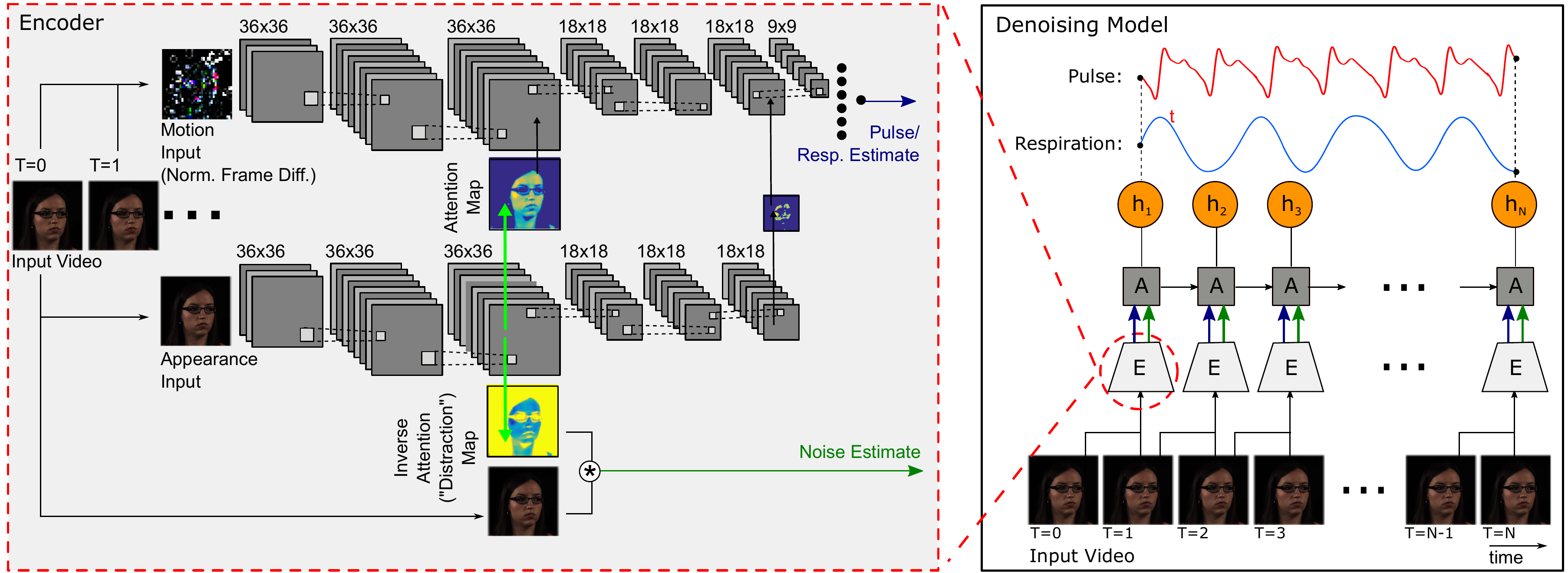}
  \caption{Proposed denoising architecture. The encoder provides the initial physiological signal and the noise estimates to the LSTM at each time step which outputs a denoised physiological signal.}
  \label{fig:architecture}
\end{figure*}

Let us take a video of a person moving as an example. The skin pixels will contain information about the physiological signal but they will also capture the body motion, as in each frame the incident light changes with the orientation of the head (see Fig.~\ref{fig:overview}). In contrast, the hair pixels will not contain information about the physiological signal (as there are no blood vessels in the hair) but will still contain information about the motion. In this section, we explain how we use those inverse attention (or ``distraction'') regions to denoise the physiological signals.  The details of the proposed deep learning architecture are provided in Fig.~\ref{fig:architecture}.

%However, because of the large variation in what constitutes non-skin regions (e.g., hair, clothing or head apparel, glasses, objects that are moving with the person or global illumination changes that impact background pixels as well as the face), the attention masks typically ignore these as they are not consistent across all videos and do not contain the signal of interest. As empirical evidence of this, prior work has shown that an attention mechanism tends to focus on skin regions containing strong physiological signals and ignores hair and background regions~\citep{chen2018deepphys} (see ``Attn. Mask'' examples in Fig.~\ref{fig:masks}). %In our example, the ignored hair regions contain some of the same noise signals as the skin regions. 
% To provide another example, variations in ambient light can often be similar across the entire frame, including the background and the skin pixels. Again, it is reasonable to think that the regions not containing physiological signals could be used to estimate the properties of the noise. We propose that using noise signals generated from an inverse attention mask will allow us to remove noise from the predicted signals of interest to obtain cleaner estimates. 

\SubSection{Physiology and Noise Encoder}
The backbone of the encoder is formed using a convolutional attention network (CAN)~\citep{chen2018deepphys}. This contains appearance and motion branches learned jointly through an attention mechanism. The appearance model is trained directly on the input video frames. It learns from the color and texture information which regions in the video are likely to contain strong physiological signals. The motion model is trained on the difference of two consecutive video frames to differentiate between the intensity variations present in the video caused by the characteristic physiological variations from those from other sources. The attention mask then reflects a heatmap of the strength of the pulsatile signal in each region of the frame. As shown in the first row of Fig.~\ref{fig:masks}, the attention masks mostly focus on skin regions known to have strong physiological signals, while ignoring other regions, such as the eyes, hair, and background regions. The CAN normally outputs a single one-dimensional (1D) physiological signal estimate. However, we perform an element-wise multiplication of the original input frame with the inverse of the attention mask weights to compute a secondary noise estimate.

We compute the noise signals at each time step by multiplying the inverse attention masks with each channel of each video frame in an element-wise manner. We then spatially average the resulting weighted pixel intensities to obtain the noise estimate:
% element-wise multiply masks and RGB video, spatially average pixel intensities
\begin{equation}
\label{eq:noise_gen}
    N_{c,t} = \frac{1}{H} \frac{1}{W} \sum\limits^{H}_{x=1} \sum\limits^{W}_{y=1}  I_{x,y,t} \circ M_{x,y,t}
\end{equation}

where $I_{t}$ and $M_{t}$ are the frame and mask at time $t$. $N_{c,t}$ is the noise estimate from each [R, G, B] camera channel $c$ at time $t$, and H and W are the image height and width, respectively. The attention and the inverse attention masks were 34 $\times$ 34 pixels and the video frames were downsampled to the same size using bicubic interpolation.

%we used the attention masks output by the CAN to guide the generation of noise estimates. 
We normalize the attention mask elements to a range between 0 and 1. To obtain a noise estimate, we set all values larger than a fixed threshold, T, to 0 and everything else to 1, creating a binary mask. Based on the experiments we found a threshold of 0.1 worked well. This binary inverse attention mask ignores regions in the video initially used to compute the physiological signals and keeps all other regions. Examples of inverse attention masks and the corresponding noise estimates are shown in the second row of Fig.~\ref{fig:masks}.

% \subsection{Noise Estimate Generation From ``Distraction" Regions}

% The signals output by the CAN are often still corrupted by noise, especially in presence of large head motion. The regions in the image which were ignored by the attention masks because of poor signal strength, may not be useful to extract the signals directly, but our hypothesis is that they contain important information about the sources of noise present in the video. 

%for each pixel at location $(h,w)$, and $M_{h,w}$ is a 1D vector of 0's and 1's obtained from the inverse attention mask for each $(h,w)$ pixel. The attention and inverse attention masks were 34 x 34 pixels and the video frames were downsampled to the same size using bicubic interpolation.

% We compared the performance with using all three [R, G, B] camera channels to only using red, green, or blue channels. We obtained comparable results with individual and all three camera channels. We report all results in this paper using all three [R, G, B] channels and include the comparison of the results with different camera channels in the Supplementary Materials.

\SubSection{Denoising Model}
% The CAN is used to generate the initial BVP estimates and the noise estimates (calculated using the inverse attention mask). 
Our denoising model is then formed as long short-term memory (LSTM) network with the encoder providing input at each time step. The goal is to learn a denoising function to further clean the physiological estimates. As input to the denoising LSTM we stacked the physiological signal and noise signal outputs generated by the encoder. 
%by the CAN with the [R, G, B] noise estimates generated with the inverse attention masks. 
The contact physiological signal (e.g., finger pulse oximeter) was used as ground truth for training. The noise estimates guide the LSTM to learn which waveform features are related to noise and which are related to the physiological signal of interest. The LSTM learns to suppress the noise from the physiological signal and outputs a cleaner waveform matching the ground truth physiological signal better (see the third row of Fig.~\ref{fig:masks}). See the video in the supplementary material for more examples of noise estimates and denoised signals. 

% \subsection{LSTM Details}
We used a two-layer bidirectional LSTM, with 128 hidden units, trained for 10 epochs with Adam optimizer~\citep{kingma2014adam} and MSE loss. Because the LSTM tends to work better on shorter sequences, we split each video to sequences of 60 samples, with 50\% overlap between time windows, which corresponded to two seconds for the 30 frames per second (fps) videos. Physiological datasets are often relatively small due to the complexity associated with collecting carefully synchronized physiological signals and high-quality videos. Therefore, we implemented the CAN and the denoising LSTM as two separate networks to reduce the number of training parameters. 
%However, the CAN signal generation and the denoising could be implemented in an end-to-end framework.

% \begin{equation}
% \label{eq:max_abs1}
%     C(t)^{'} = \frac{C(t)^{'}}{\textrm{max}(|C(t)^{'}|)}
% \end{equation}

% \begin{equation}
% \label{eq:max_abs2}
%     N_{C}^{'}(t) = \frac{N_{C}^{'}(t)}{\textrm{max}(|N_{C}^{'}(t)|)}
% \end{equation}

% We implemented the CAN and the denoising LSTM as two separate networks (A BETTER WAY OF DESCRIBING WHY YOU DID IT THIS WAY IS THAT TWO SEPARATE NETWORKS IS FEWER PARAMETERS THAN A 3D CNN) because the CAN network uses two consecutive frames only to compute the signals, whereas our denoising LSTM uses longer sequences to learn the waveform dynamics and to remove noise. However, the CAN signal generation and the denoising could be implemented in an end-to-end framework by changing the architecture to use an input of the same length. 

\Section{Datasets}
\label{sec:datasets}

%\begin{figure*}[h!]
\begin{wrapfigure}{L}{0.5\textwidth}
\includegraphics[width=0.5\textwidth]{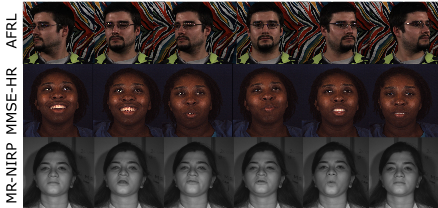}
\caption{Examples of images used to evaluate our proposed approach.}
\label{fig:datasets}
%\vspace{-0.5cm}
\end{wrapfigure}
%\end{figure*}

We evaluated our approach on the following two RGB video datasets and an NIR video dataset. %For extended descriptions of the data see the supp. material.

% The videos feature male and female participants with diverse skin tones, some with facial hair and/or glasses, a variety of body motions, including large out of plane head rotation, facial expressions and translational head motion. More details about the datasets can be found in the supplementary materials.

\textbf{AFRL}~\citep{estepp2014recovering}: 300 videos of 25 participants were recorded as 658 $\! \times \!$ 492 pixel images at 120 fps. Fingertip reflectance photoplethysmograms (PPG), electrocardiograms (ECG), and respiratory signals were recorded as ground truth signals. We used the ECG signals to compute the HR estimation errors, the PPG signals to train the network for estimating HR, and respiratory signals for computing the errors and training the network for BR estimation. Each participant was recorded 12 times in each five minute experiment with varying motion and two different backgrounds. The participants: 1) sat still and rested their chin on a headrest, 2) sat still without the headrests, 3) moved their head horizontally at a speed of 10 degrees/second, 4) 20 degrees/second, 5) 30 degrees/second, 6) reoriented their head randomly once every second. We center-cropped the ARFL video frames to 492 $\! \times \!$ 492 pixels to remove the blank background areas. 
% We removed the first and last 15 seconds of each 5.5 minute recording.  

% The participants were asked to sit still for the first two experiments, and in the subsequent four motion experiments, the participants were asked to move their head horizontally from left to right with an increasing speed.

\textbf{MMSE-HR}~\citep{zhang2016multimodal}: 102 videos of 40 participants were recorded at 25 fps capturing 1040 $\! \times \!$ 1392 resolution images during spontaneous emotion elicitation experiments. Ground truth blood pressure (BP) wave was measured at 1000 fps and average HR updated after every heart beat. We used the blood pressure waves to train the network and the average HR to compute the HR estimation errors. 19 videos had erroneous average HR estimates, so we recomputed them using the BP wave.

% The MMSE-HR dataset is more challenging than the AFRL dataset because it contains uncontrolled and non-rigid face motion caused by emotion elicitation experiments, as opposed to the more controlled and rigid head motion present in AFRL.

% We removed 15 videos which had erroneous ground truth heart rate measurements~\citep{macwan2019heart}. The list of the removed videos is provided in the Supplementary Materials.

\textbf{MR-NIRP (NIR)}~\citep{nowara2018sparseppg}: Eight participants were recorded with a NIR camera. The videos were recorded at 640$ \! \times \!$ 640 resolution and 30 fps. Fingertip transmission photoplethysmograms were recorded as ground truth signals. Each participant was recorded twice, once sitting still and once performing motion tasks involving talking and randomly moving the head. Because the background in MR-NIRP was not uniform, we applied face detection in the first video frame and cropped a rectangular region with 110\% width and height of the detected bounding box.

\Section{Experiments}
\label{sec:Experiments}

\textbf{Training the Encoder.} Due to the large number of parameters we pretrain the encoder on the largest dataset (AFRL~\citep{estepp2014recovering}) and lock the weights. When training the encoder the loss is calculated as the mean squared error between the physiological estimate and the ground truth. We do not compute the loss using the noise estimate as there is no ground truth noise signal. In our experiments we performed training and testing separately for each of six motion tasks from the AFRL dataset with a participant-independent cross-validation, leaving out 20\% of the participants in each validation split. For experiments on the MMSE-HR and MR-NIRP datasets we used the trained model from Task 2 as these contained the most similar amplitude motions. To maximize the diversity of the participants that this model was trained on to improve the generalizability to new datasets, we instead used a subject-dependent cross-validation, using four minutes of each video for training and one minute for testing. 

\textbf{Training the Denoising Model.} When evaluating on the AFRL dataset we trained the denoising model with the same subject-independent procedure as for the encoder on AFRL. The MMSE-HR dataset has fewer videos than the AFRL dataset; therefore, we used a leave-one-subject-out cross-validation where we left out all videos of one subject and trained the model on all remaining videos, repeating this for each subject. The MR-NIRP dataset was small and not suited for training the networks, so we used the LSTM trained on the AFRL dataset. This allowed us to test cross-dataset generalization. 

%\subsection{Evaluation Metrics}
% \textbf{Signal Processing and Normalization.} 
% We detrended signals (eq.~\ref{eq:detrend1}) 
% \begin{equation}
% \label{eq:detrend2}
%     N_{C}(t) = (I - (I + \lambda^{2} D_{2}^{T} D_{2})^{-1}) N_{C}(t)  
% \end{equation}
We bandpass filtered ([0.7 Hz, 2.5 Hz]) and detrended the signals~\citep{tarvainen2002advanced}. We normalized the signals by subtracting the temporal mean, dividing by the temporal standard deviation in each video, and normalized their amplitudes to -1 and 1. We resampled all sequences to 30 fps. The signals from each video were divided into 30-second non-overlapping windows. We evaluated the performance of our proposed denoising approach using mean absolute error (MAE), root mean square error (RMSE), Pearson's correlation coefficient ($\rho$) between the estimated HR and the ground truth HR, SNR of the estimated physiological signals~\citep{CHROMdeHaan}, and waveform mean absolute error (WMAE) computed between the estimated and the ground truth signal. See the supplementary material for the definitions of the error metrics.

% \begin{equation}
% \label{eq:detrend1}
%     C(t) = (I - (I + \lambda^{2} D_{2}^{T} D_{2})^{-1}) C(t)  
% \end{equation}
% where $I$ is the identity matrix and $D_{2}$ is the second order difference matrix defined in~\citep{tarvainen2002advanced}.

% \begin{equation}
% \label{eq:norm1}
%     C(t)^{'} = \frac{C(t) - \mu_{C}}{\sigma_{C}}
% \end{equation}

% where $\mu$ is the pixel mean and $\sigma$ is the standard deviation.

% \dan{We need to add the filter steps - they should have been the same for all methods but I don't think there is anywhere in the paper that we mention the bandpass filter order or cut-offs} 

% % and 0.08 and 0.5 Hz for BR

% \subsection{Denoising Framework}

% \begin{figure*}[t!]
%   \includegraphics[width=0.4\textwidth]{blank.pdf}
%   \caption{BVP and BVP second order derivatives measured using: blue) Ours (w/ LSTM), red) CAN~\citep{chen2018deepphys}. The second order derivative of the waveform reveals the diastolic peak (or inflection).}
%   \label{fig:waveform_dynamics}
% \end{figure*}

\Section{Results and Discussion}
\label{sec:results}

% Our proposed approach uses an inverse attention mask to generate noise estimates and increase the signal-to-noise ratio of pulse estimates recovered from videos. Our assumption being that the attention masks give weight to pixels that contain the strongest pulse signal but do not heavily weigh pixels that contain noise information that could be used to boost the pulse SNR. 
We compare four variants of our proposed approach to four state-of-the-art methods for recovering the pulse signal~\citep{chen2018deepphys,poh2010non,CHROMdeHaan,wang2017algorithmic} and two methods for recovering the breathing signal~\citep{chen2018deepphys,tarassenko2014non} (see the supplementary material for implementation and signal pre-processing details). The variants of our approach we compare are: training an our model with noise estimates as input (`Distraction") and without noise estimates as input (``No Noise") 
%to denoise the signals to simply using an LSTM without the noise estimates to filter the signals ,
and subtracting the noise estimates from the physiological signal either in the frequency domain - ``Freq. Sub." - or time domain - ``Wave. Sub.". %\textbf{Bold} numbers reflect the best performance in all tables.

%it also introduces new artifacts in the recovered signal leading to erroneous waveform dynamics and even larger errors in heart rate estimation, compared to training an LSTM network to denoise the BVP signals. 
% \textbf{How does our approach compare to existing methods?}
\textbf{Heart Rate Estimation.} Our method achieves lower HR MAE, RMSE and waveform MAE and higher HR correlation ($\rho$) and SNR (see Tables~\ref{tab:HR_BR}) compared to previous approaches on two large datasets. On the AFRL dataset the MAE is reduced from 2.93 beats per minute (BPM) to 2.25 BPM (25\% reduction in error), and on the MMSE-HR dataset the MAE is reduced from 3.74 BPM to 2.50 BPM (33\% reduction in error). This shows that information excluded by the attention mask can be successfully leveraged to remove noise, leading to substantial improvements in signal quality. Moreover, the proposed denoising approach is able to recover the subtle waveform dynamics, reducing the waveform MAE by more than 50\% on MMSE-HR. While simply subtracting the noise from the signals in the frequency domain often improved the SNR, it did not improve the heart rate estimates. Subtracting the noise signal in the time domain performed even worse and had a particularly negative impact on the BVP SNR. All results were statistically significant (p $<$ 0.01) -- see supplementary material for F-test results.

\begin{table*}[hbt]
	\caption{Heart rate (HR) and breathing rate (BR) estimation on 3 public datasets. Including the ``distraction" regions significantly improves the HR and BR estimation.}
	\label{tab:HR_BR}
	\centering
	\scriptsize
	\setlength\tabcolsep{0.8pt} % default value: 6pt
	\begin{tabular}{r|ccccc|ccccc|ccccc||ccccc}
	%\toprule
	& \multicolumn{15}{c}{\textbf{Heart Rate}} &  \multicolumn{5}{c}{\textbf{Breathing Rate}} \\
	\hline
	& \multicolumn{5}{c}{\textbf{AFRL}} & \multicolumn{5}{c}{\textbf{MMSE-HR}} & \multicolumn{5}{c}{\textbf{MR-NIRP(NIR)}} & \multicolumn{5}{c}{\textbf{AFRL}} \\ 
	
         \textbf{Method} & MAE & RMSE  & SNR & $\rho$ & WMAE & MAE & RMSE  & SNR & $\rho$ & WMAE & MAE & RMSE  & SNR & $\rho$ & WMAE & MAE & RMSE  & SNR & $\rho$ & WMAE \\ \hline \hline
          Distraction  & 2.25 & 5.68 & 6.44 & 0.87 & 0.21 & \textbf{2.27} & \textbf{4.90} & \textbf{5.00} & \textbf{0.94} & \textbf{0.19} & \textbf{2.34} & \textbf{4.46} & 2.27 & \textbf{0.85} & 0.45 & \textbf{2.44} & \textbf{4.23} & \textbf{14.20} & \textbf{0.35} & 0.28 \\
          No Noise & \textbf{2.12} & \textbf{5.37} & \textbf{6.86} & \textbf{0.88} & \textbf{0.21} & 2.80 & 6.36 & 4.30 & 0.90 & 0.21 & 2.56 & 5.23 & \textbf{2.28} & 0.80 & 0.40 & 2.49 & 4.26 & 14.06 & 0.34 & \textbf{0.27} \\
          Freq. Sub. & 2.92 & 6.67 & 3.66 & 0.82 & 0.24 & 3.97 & 9.93 & 4.49 & 0.76 & 0.57 & 8.58 & 17.59 & -4.56 & -0.11 & \textbf{0.31} & 5.03 & 7.45 & 7.78 & 0.12 & 0.31 \\ 
          Wave. Sub. & 2.92 & 6.66 & 3.09 & 0.82 & 0.24 & 6.09 & 10.84 & -4.75 & 0.71 & 0.55 & 8.83 & 17.00 & -4.69 & -0.17 & \textbf{0.31} & 7.21 & 8.83 & 4.12 & -0.03 & 0.37 \\
          \hline
        CAN  & 2.93 & 6.69 & 3.36 & 0.82 & 0.23 & 4.06 & 9.51 & 0.63 & 0.77 & 0.57 & 7.78 & 16.81 & -3.24 & -0.03 & 0.36 & 4.86 & 7.32 & 8.33 & 0.10 & 0.27 \\
        POS & 4.36 & 9.45 & 0.73 & 0.74 & 0.45 & 3.90 & 9.61 & 2.33 & 0.78 & 0.39 & -- & --  & -- & --  & -- & -- &  -- &  -- & -- &  --  \\
        CHROM & 4.07 & 9.72 & 0.29 & 0.72 & 0.41 & 3.74 & 8.11 & 1.90 & 0.82 & 0.37 & -- & --  & -- & --  & -- & -- &  --  & --  & -- &  -- \\
        ICA & 5.78 & 11.82 & 0.42 & 0.58 & 0.43 & 5.44 & 12.00 & 3.03 & 0.66 & 0.42 & -- & --  & -- & --  & -- & -- & -- & -- &  -- & --\\
        Tar.  & -- & --  & -- & --  & --  & -- & -- & -- & -- & -- &  -- & -- & -- & -- & -- &  3.68 & 5.52 & -6.22 & 0.29  & 0.29 \\
        %%\bottomrule
  \end{tabular}
%   \\ MAE = Mean Absolute Error in HR estimation [BPM], RMSE = Root Mean Squared Error in HR estimation [BPM], SNR = BVP Signal-to-Noise Ratio [dB], $\rho$ = Pearson Correlation in HR estimation, WMAE = Waveform MAE, BR MAE = Mean Absolute Error in Breathing Rate Estimation, SNR BR = Breathing Signal-to-Noise Ratio [dB].\\
CAN =~\citep{chen2018deepphys}, POS =~\citep{wang2017algorithmic}, CHROM =~\citep{CHROMdeHaan}, ICA =~\citep{poh2010non}, Tar =~\citep{tarassenko2014non}.
\end{table*}

\begin{wrapfigure}{L}{0.5\textwidth}
  \includegraphics[width=0.5\textwidth]{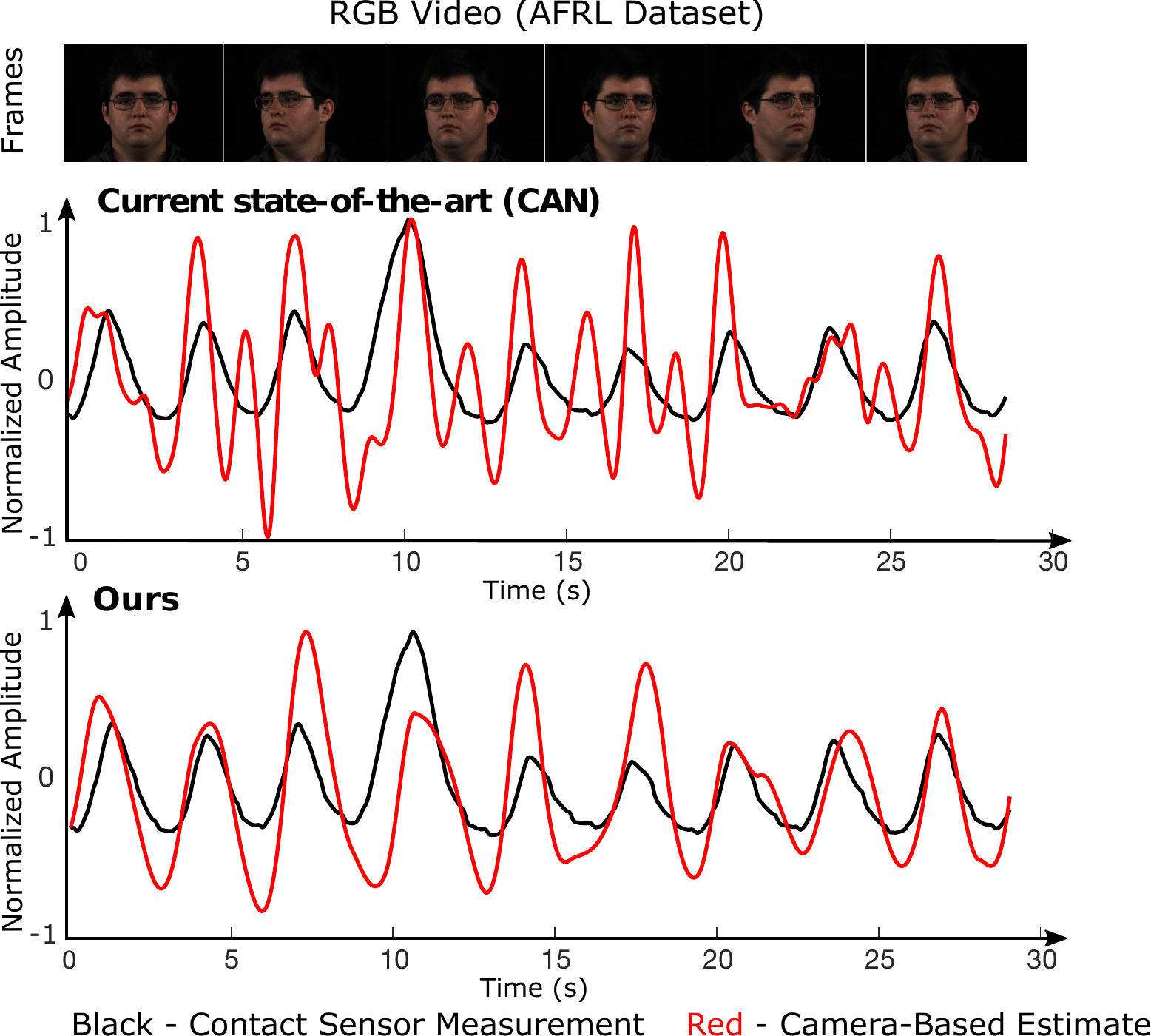}
  \caption{Respiration signals output by a state-of-the-art network and our denoising method. The signals produced by our method are cleaner and match the ground truth better.} 
  \label{fig:BR}
  \vspace{-0.5cm}
\end{wrapfigure}

\textbf{Breathing Rate Estimation.} In addition to estimating heart rate, which is based on intensity variations in the skin, our method can also be used to estimate breathing rate (BR) which is based on motion variations and is more challenging in presence of body motions. Only the AFRL dataset~\citep{estepp2014recovering} had gold standard reference breathing signals, therefore we were not able to evaluate our BR results on the other datasets. Our method achieves a reduction in MAE from 3.68 BPM to 2.44 BPM (a 34\% error reduction) over the baselines and an increase in SNR of 5.87 dB (Table~\ref{tab:HR_BR}). Our method also obtains cleaner breathing signals compared to the baseline (Fig.~\ref{fig:BR}).

\begin{wraptable}{L}{0.6\textwidth}
	\caption{Training on AFRL Task 1 and testing on Task 6. The ignored regions help when the training and test set are very different.}
	\label{tab:no_noise}
	\centering
	\scriptsize
	\setlength\tabcolsep{0.65pt} % default value: 6pt
	\begin{tabular}{rc|ccccc|ccccc}
	%\toprule
	&& \multicolumn{5}{c}{\textbf{Heart Rate}} & \multicolumn{5}{|c}{\textbf{Breathing Rate}} \\
        \textbf{Method} && MAE & RMSE  & SNR & $\rho$ & WMAE &  MAE & RMSE  & SNR & $\rho$ & WMAE \\ \hline \hline
        Distraction  && \textbf{5.29} & \textbf{9.33} & -2.07 & \textbf{0.70} & 0.32 & \textbf{4.28} & \textbf{6.00} & 5.93 & \textbf{0.10} & 0.34 \\
        No Noise && 5.61 & 9.72 & \textbf{-1.91} & 0.67 & 0.32 & 4.38 & 6.15 & \textbf{5.96} & 0.07 & 0.34\\
        %CAN && 5.34 & 9.75 & -2.23 & 0.64 & \textbf{0.30} \\ 
        %\bottomrule
  \end{tabular}
\end{wraptable}

\textbf{True Benefit of Distraction Regions.} Using our model without noise estimates from the output of the CAN to the LSTM works well when the signals do not change much over time, and when the noise in the training and test sets is similar, e.g., training and testing on AFRL (Table~\ref{tab:HR_BR}). However, including the distraction regions yields improvements in both HR and BR estimates when the signal varies over time or there is a large domain gap between training and testing sets. For example, distraction regions improve the performance on MMSE-HR which has sudden pulse variations, uncontrolled motion, and the presence of facial expressions and on the more challenging NIR MR-NRIP dataset (Table~\ref{tab:HR_BR}). Moreover, including the distraction regions improves the HR and BR estimation accuracy when we train our model only on stationary videos of AFRL (Task 1) and test on videos with large random motions (Task 6) (Table~\ref{tab:no_noise}). SNR is often higher in the ``no noise'' condition because it simply produces a smoother signal leading to greater sparsity in the frequency domain. However, the dominant frequency of the signal (used to compute HR and BR) is often erroneous, resulting in higher MAE and RMSE, and lower $\rho$. These results show that the distraction signal is useful above and beyond including a temporal component to the model.

\textbf{Transfer learning.} NIR videos of MR-NIRP are more challenging than RGB because the physiological signal is an order of magnitude weaker in the NIR range compared to the visible range, making it very prone to motion artifacts. When trained solely on RGB videos (AFRL dataset) without any fine-tuning, our method outperforms all the baselines across all five metrics on the NIR videos from the MR-NIRP dataset. As shown in Table~\ref{tab:HR_BR} the MAE drops from 7.78 BPM to 2.34 BPM (70\% reduction in error). Other baseline methods require multiple color channels and therefore cannot be applied to NIR videos. %Our denoising approach could potentially generalize across the electromagnetic spectrum, including thermal imaging in far-infrared. However, to our knowledge there are no publicly available thermal video datasets with ground truth physiological measurements. 

% AFRL We achieve MAE [BPM] on all motion tasks in AFRL of: Ours (LSTM with noise) = 2.25, LSTM without noise = \textbf{2.12}, CAN = 2.93. 
\textbf{Varying Head Motion.} We performed an analysis of the performance on motion tasks of AFRL~\citep{estepp2014recovering} (see Tables~\ref{tab:motion}). Our method shows improvements over the baseline methods on videos across all head motions. For instance, on videos with an angular head rotation of 30 deg/sec (Task 4) the HR MAE was reduced from 2.82 BPM to 1.94 BPM (30\% reduction in error) and BR MAE was reduced from 4.85 BPM to 2.88 BPM (40\% reduction in error). %In videos with greater motion, the motion artifacts affect the signal both inside and outside of the attention mask more and therefore the noise cancellation offered by our approach has the greatest effect. 
\def\methodsa{\makecell[l]{de Haan et al.~\citep{CHROMdeHaan}, \\ McDuff et al.~\citep{mcduff20145band}, \\
Poh et al.~\citep{poh2010non}, \\ Wang et al.~\citep{wang2017algorithmic}, \\Estepp et al.~\citep{estepp2014recovering}}}

\begin{wraptable}{L}{0.6\textwidth}
	\caption{Motion increasing from 1 to 6 on AFRL}
	\label{tab:motion}
	\centering
    \scriptsize
    \setlength\tabcolsep{0.8pt} % default value: 6pt
	\begin{tabular}{r|cccccc|cccccc}
	%\toprule
	& \multicolumn{6}{c}{\textbf{Heart Rate MAE}} & \multicolumn{6}{|c}{\textbf{Breathing Rate MAE}} \\
	\hline
	\textbf{Method} & 1 & 2 & 3 & 4 & 5 & 6 & 1 & 2 & 3 & 4 & 5 & 6 \\ \hline \hline
% 	\makecell[r]{Ours, \\ (LSTM)}
       Distraction &  \textbf{1.06} & 2.11 & \textbf{1.79} & \textbf{1.94} & 2.50 & 4.78 & \textbf{1.42} & \textbf{1.86} & 1.88 & \textbf{2.88} & \textbf{2.87} & \textbf{4.15} \\
    %  \makecell[r]{Ours, \\ (LSTM, \\ No Noise)} \\
      No Noise & 1.14 & 1.90 & 1.80 & 3.39 & \textbf{2.04} & \textbf{4.52} & 1.47 & 1.95 & \textbf{1.68} & 2.96 & 2.99 & \textbf{4.15}  \\ 
      Freq. Sub. & 1.52 & 2.62 & 2.51 & 3.00 & 2.58 & 5.30 & 4.30 & 5.35 & 4.89 & 5.27 & 5.09 & 5.26  \\
      Wave. Sub. & 1.57 & 2.59 & 2.53 & 3.03 & 2.72 & 5.09 & 4.31 & 5.24 & 4.88 & 5.17 & 5.08 & 5.19 \\
      \hline
    CAN &  1.52 & 2.61 & 2.51 & 3.00 & 2.62 & 5.34 & 4.24 & 5.17 & 4.58 & 5.09 & 4.92 & 5.15  \\
    POS &  1.42 & \textbf{1.52} & 2.84 & 3.86 & 6.33 & 10.16 & -- & -- & -- & -- & -- & --\\ 
    CHROM & 1.33 & 1.62 & 2.87 & 2.82 & 3.91 & 11.86 & -- & -- & -- & -- & -- & --\\ 
    ICA & 2.18 & 2.64 & 4.74 & 4.93 & 7.02 & 13.18 & -- & -- & -- & -- & -- & -- \\ 
    Tar. & -- & -- & -- & -- & -- & -- & 2.51 & 2.53 & 3.19 & 4.85 & 4.22 & 4.78 \\
    %\bottomrule
      \end{tabular}
\end{wraptable}

\textbf{Inverse Mask Definition.} We tested computing the inverse attention mask in two different ways. The first, as a matrix of continuous values in which each element of the inverse mask $M$, $M_{i,j}$, was 1 - $A_{i,j}$ where $A$ is the attention mask weights normalized from 0 to 1. The second approach was to threshold these values to create a binary mask where $A'_{i,j}$ = 1, if $A_{i,j}>$T. Where T is a threshold from 0 to 1. We found we obtain comparable results with binary (2.25 BPM) or continuous inverse attention masks (2.10 BPM). We also found that the results were not very sensitive to the value of T (see supplementary materials).

\textbf{Importance of Different Distraction Regions.} Certain regions in the video may contain more useful information about the sources of noise than others. For example, regions closer to the face may contain more information about the motion of the participant, while regions farther in the background may contain more information about other sources of noise, such as illumination changes. We compared separately using noise estimates from distraction regions closer to the face (center of the frames) and further from the face (edges of the frames). When motion was small, all regions contributed similarly to denoising (MAE = 1.08 BPM with center regions and MAE = 1.07 BPM with edges). But when there was large head motion, regions close to the head (center of the frames) helped the most (MAE = \textbf{6.53} BPM with center regions and MAE = 8.74 BPM with edges). See supplementary materials for detailed results.

% If the attention mask is poorly estimated, the method won't work well. But so long as the signal of interest is stronger in the attention mask than in the inverse attention mask, our method should show some benefit. 

\textbf{Effect of Glasses.} Interestingly, our method performed best on subjects who wore glasses. The attention masks for subjects with and without glasses were comparably good. However, CAN performed worse on subjects with glasses and our approach offered a large improvement on those videos (MAE [BPM] with glasses: Ours = \textbf{2.17}, CAN = 3.33, and without glasses: Ours = \textbf{2.55}, CAN = 2.57). See supplementary materials for example attention masks and results.

\Section{Conclusion}
We have presented a novel approach for generating noise estimates from inverse attention masks to improve camera-based physiological signal measurements. We hypothesized that the noise affecting regions used by the attention masks to compute the signal of interest would likely be present in other regions in the video which are ignored by the attention masks. Our proposed denoising method outperformed all state-of-the-art methods in heart rate and breathing rate estimation from videos. The recovered BVP signals are also sufficient to recover subtle waveform dynamics present in the ground truth contact signals, including the dicrotic notch and the diastolic peak. Our approach trained on RGB videos showed strong cross-dataset and cross-modality generalizability, outperforming the existing methods on challenging NIR videos.

% We evaluated our approach on the task of non-contact physiological measurement which has many applications (infant monitoring, critical care, health tracking, affect measurement). All these applications benefit from accurate measurement, especially in the presence of noise (e.g., motion, ambient illumination changes) as is often the case in realistic unconstrained scenarios. 

% To test this hypothesis we used a convolutional attention network architecture to compute the BVP signals from video frames and learn a spatial attention mask. We then introduced an inverse attention mask to compute the noise estimates and a denoising mapping to remove the estimated noise from the BVP signals. 

%\subsubsection*{Acknowledgments}
\bibliography{egbib}
\bibliographystyle{2021_conference}

\appendix 

\section{Appendix}

\subsection{Evaluation Metrics}
To evaluate the performance of our proposed approach we used the following four standard error measures (MAE, RMSE, Correlation, SNR), and we defined a new measure (Waveform MAE) to measure the waveform dynamics.   

\textbf{Mean absolute error (MAE)}:
\begin{equation}
    \text{MAE} = \frac{\sum\limits^N_{i=1} |R_{i} - \widehat{R}_{i}|}{N}
\end{equation}

where $N$ is the total number of time windows, $R_{i}$ is the ground truth heart rate (HR) measured with a contact sensor for each 30 second time window and $\widehat{R}_{i}$ is the estimated HR from the video.

\textbf{Root Mean Square Error (RMSE)}:
\begin{equation}
    \text{RMSE} = \sqrt{\frac{\sum\limits^N_{i=1} (R_{i} - \widehat{R}_{i})^{2}}{N}}
\end{equation}

\textbf{Pearson’s Correlation Coefficient ($\rho$)}: computed between HR estimates from each time window $\widehat{R} = [{\widehat{R})(1), ..., \widehat{R}(N)}]$ and the ground truth HR measurements $R = [{R(1), ..., R(N)}]$.

\textbf{Signal-to-noise ratio (SNR):} calculated as the ratio of the area under the curve of the power spectrum around the first and the second harmonic of the ground truth HR frequency divided by the rest of the power spectrum within the physiological range of 42 to 240 bpm  \cite{CHROMdeHaan}: 

\begin{equation}
   \text{SNR} = 10\log_{10}\bigg( \frac{\sum\limits^{240}_{42}((U_{t}(f)S(f))^{2}}{\sum\limits^{240}_{42}((1 - U_{t}(f))S(f))^{2}}\bigg)
\end{equation}

where $S$ is the power spectrum of the estimated iPPG signal, $f$ is the frequency in beats per minute (BPM) and $U_{t}(f)$ is equal to one for frequencies around the first and second harmonic of the ground truth HR (HR-6 bpm to HR+6 bpm and 2*HR-6 bpm to 2*HR+6 bpm), and 0 everywhere else.

\textbf{Waveform Mean Absolute Error (WMAE):}
\begin{equation}
    \text{WMAE} = \frac{\sum\limits^N_{i=1} |W_{i} - \widehat{W}_{i}|}{N}
\end{equation}
where $W_{i}$ is the ground truth pulse waveform obtained with the contact sensor for each 30 second time window and $\widehat{W}_{i}$ is the estimated pulse waveform from the video.

\subsection{Baseline Methods}
We compared the performance of our proposed approach to state-of-the-art supervised method using a convolutional attention network (CAN) and three unsupervised methods described below.

For the CHROM, ICA and POS methods face detection was first performed using MATLAB's face detection (\texttt{vision.CascadeObjectDetector()}). This was fixed for all methods, to avoid the influence of the face detector on performance. For the CAN method following the implementation in~\citep{chen2018deepphys} we did not use face detection but rather we passed the full frame to the network after cropping the center portion to make the frame a square with W=H.

\textbf{CHROM}~\citep{CHROMdeHaan}. This method uses a linear combination of the chrominance signals obtained from the RGB video.  The [\begin{math}x_R\end{math}, \begin{math}x_G\end{math}, \begin{math}x_B\end{math}] signals are filtered using a zero-phase, 3rd-order Butterworth bandpass filter with pass-band frequencies of [0.7 2.5] Hz. Following this, a moving window method of length 1.6 seconds (with overlapping windows and a step size of 0.8 seconds) is applied. Within each window the color signals are normalized by dividing by their mean value to give [\begin{math}\bar{x_r}\end{math}, \begin{math}\bar{x_g}\end{math}, \begin{math}\bar{x_b}\end{math}]. These signals are bandpass filtered using zero-phase forward and reverse 3rd-order Butterworth filters with pass-band frequencies of [0.7 2.5] Hz. The filtered signals [y$_{r}$, y$_{g}$, y$_{b}$] are then used to calculate \textit{S$_{win}$}:

\begin{equation}
S_{win} = 3( 1 - \frac{\alpha}{2} )y_{r} - 2(1 + \frac{\alpha}{2})y_{g} + \frac{3\alpha}{2}y_{b}
\end{equation}

Where $\alpha$ is the ratio of the standard deviations of the filtered versions of A and B:
\begin{equation}
    A = 3y_r - 2y_g
\end{equation}
\begin{equation}
B = 1.5y_r + y_g - 1.5y_b
\end{equation}
The resulting outputs are scaled using a Hanning Window and summed with the subsequent window (with 50\% overlap) to construct the final blood volume pulse (BVP) signal.

\textbf{ICA}~\citep{poh2010non}. This approach involves spatial averaging the pixels by color channel in the region of interest (ROI) for each frame to form time varying signals [x$_{R}$, x$_{G}$, x$_{B}$]. Following this, the
observation signals are detrended. A
Z-transform is applied to each of the detrended signals. The Independent Component Analysis (ICA) (JADE implementation) is applied to the normalized color signals. 

\textbf{POS}~\citep{wang2017algorithmic}. The intensity signals [\begin{math}x_R\end{math}, \begin{math}x_G\end{math}, \begin{math}x_B\end{math}] are computed. A moving window of length 1.6 seconds (with overlapping windows and with a step size of one frame), is applied. For each time window, the signal is divided by its mean to give [\begin{math}\bar{x_r}\end{math}, \begin{math}\bar{x_g}\end{math}, \begin{math}\bar{x_b}\end{math}]. Following this, \textit{X$_s$} and \textit{Y$_s$} are calculated where:
\begin{equation}
    X_s = \bar{x}_g - \bar{x}_b
\end{equation}
\begin{equation}
    Y_s = -2\bar{x}_r + \bar{x}_g + \bar{x}_b
\end{equation}
\textit{X$_s$} and \textit{Y$_s$} are then used to calculate \textit{S}$_{win}$, where:
\begin{equation}
    S_{win} = X_s + \frac{\sigma(X_s)}{\sigma(Y_s)}Y_s
\end{equation}
The resulting outputs of the window-based analysis are used to construct the final BVP signal in an overlap add fashion.

\textbf{CAN}~\citep{chen2018deepphys} Supervised convolutional attention neural network described in detail in the main text~\citep{chen2018deepphys}. Following the implementation in that paper we did not use face detection but rather we pass the full frame to the network after cropping the center portion to make the frame a square with W=H.

\textbf{Signal Pre-processing.} We bandpass filtered the physiological signals and noise estimates to 0.7 Hz - 2.5 Hz range and detrended them~\citep{tarvainen2002advanced} before feeding them into the LSTM. We set the detrending parameter $\lambda$ for each dataset based on the video frame rate ($\lambda$ = 500 for AFRL~\citep{estepp2014recovering} and $\lambda$ = 50 for MMSE-HR~\citep{zhang2016multimodal} and MR-NIRP~\citep{nowara2018sparseppg}.). We normalized the signals and noise estimates with AC/DC normalization by subtracting the temporal mean and dividing by the temporal standard deviation computed for each video. We additionally normalized the amplitude range of the signals, noise estimates and the ground truth signals to -1 and 1. Finally, we resampled all sequences to 30 fps.

\textbf{Statistical Significance.} We computed F-tests to verify that our errors had significantly lower variance (spread) than the baselines. For AFRL and MR-NIRP which had longer videos, we computed the error metrics for each video, and for the shorter MMSE-HR, we computed them for all time windows in the dataset. In addition to lower mean errors, for all datasets our approach led to significantly lower spread in the MAE and RMSE. AFRL (300 videos): MAE: F = 0.54, p $<$ 0.01, RMSE: F = 0.56, p $<$ 0.01, MMSE-HR (131 windows): MAE: F = 0.26, p $<$ 0.01, RMSE: F = 3.92 p $<$ 0.01, MR-NIRP (15 videos): MAE  F = 7.94 p $<$ 0.01, RMSE F = 6.63, p $<$ 0.01.

\subsection{Comparison of Noise Estimation}
\begin{wraptable}{L}{0.6\textwidth}
	\caption{Participant independent performance of pulse measurement on AFRL~\citep{estepp2014recovering}. There was no systematic benefit of using R, G, B or RGB inputs or using the binary vs. continuous mask. We used the binary mask with RGB inputs for the results shown in the main paper.}
	\label{tab:noise_channels_mean}
	\centering
	\scriptsize
	\setlength\tabcolsep{3pt} % default value: 6pt
	\begin{tabular}{rc|ccccc}
	%\toprule
		& \multicolumn{6}{c}{\textbf{AFRL (All Tasks)~\citep{estepp2014recovering} }} \\
        \textbf{Method} && MAE & RMSE  & SNR & $\rho$ & WMAE \\ \hline \hline
        Ours (LSTM RGB binary mask) && 2.25 & 5.68 & 6.44 & 0.87 & 0.21  \\ 
        Ours (LSTM Red Binary Mask) && 2.09 & 5.19 & 6.70 & \textbf{0.89} & 0.21 \\
        Ours (LSTM Green Binary Mask) && \textbf{2.04} & \textbf{5.11} & \textbf{6.84} & \textbf{0.89} & 0.21 \\
        Ours (LSTM Blue Binary Mask) &&  2.18 & 5.27 & 6.59 & 0.88 & 0.21  \\
        Ours (LSTM RGB Continuous Mask) &&  2.10 & 5.61 & \textbf{7.11} & 0.87 & \textbf{0.20} \\
        
        %\bottomrule
   \end{tabular}
   \tiny
\end{wraptable}
\textbf{Noise Signal Definition.} We compared the performance of our proposed denoising framework with noise channels computed from a single red, green or blue camera channel to using all three R, G, B channels. We hypothesized that the blue channel might be the best one for the noise representation for the physiological signals because the hemoglobin present in blood has the lowest absorption in the blue light spectrum and its intensity variations would be least related to blood flow. Conversely, the green channel could also be a useful noise representation, because it would contain information most similar to the physiological signals since the hemoglobin has the largest absorption in the green spectrum. However, we found that there is not a large difference between using any one of the single channels or all three channels. We report the detailed results in Table~\ref{tab:noise_channels_mean} on the AFRL dataset~\citep{estepp2014recovering}.

\textbf{Inverse Mask Definition.} We also compared computing noise using a binary and a continuous inverse attention mask. The continuous mask was computed as a matrix of continuous values in which each element of the inverse mask $M$, $M_{i,j}$, was 1 - $A_{i,j}$ where $A$ is the attention mask weights normalized from 0 to 1. The binary mask was computed by thresholding these values, where $A'_{i,j}$ = 1, if $A_{i,j}>$T, where T is a threshold from 0 to 1. We found that we obtained comparable results with the binary and continuous masks as shown in Table~\ref{tab:noise_channels_mean}.

\begin{table}
	\caption{Different Distraction Regions on AFRL~\citep{estepp2014recovering}}
	\label{tab:dif_ROIs}
	\centering
    \scriptsize
    \setlength\tabcolsep{3pt} % default value: 6pt
	\begin{tabular}{rc|cccccc|cccccc}
	%\toprule
    && \multicolumn{6}{c}{\textbf{MAE}} & \multicolumn{6}{c}{\textbf{BVP SNR}} \\
	\textbf{Method} && 1 & 2 & 3 & 4 & 5 & 6 & 1 & 2 & 3 & 4 & 5 & 6  \\ \hline \hline
    Edges && \textbf{1.07} & \textbf{2.10} & 1.92 & 2.10 & 2.68 & 8.74 & \textbf{10.52} & 7.23 & 8.59 & 6.04 & 3.07 & -5.83 \\
    Center &&  1.08 & 2.11 & \textbf{1.75} & \textbf{2.00} & \textbf{2.43} & \textbf{6.53} & 10.50 & \textbf{7.28} & \textbf{8.72} & \textbf{6.33} & \textbf{3.89} & \textbf{-4.47} \\

    %\bottomrule
      \end{tabular}
\end{table}
\textbf{Different Distraction Regions.} We compared separately using noise estimates from distraction regions closer to the face (``Center" of the frames) and further from the face (``Edges" of the frames). We used an LSTM model trained on all ignored regions for this experiment. When motion was small, all regions contributed similarly to denoising. But when there was large head motion, regions close to the head (center of the frames) helped the most. See Table~\ref{tab:dif_ROIs}.

\begin{wraptable}{L}{0.5\textwidth}
	\caption{Effect of Glasses on AFRL~\citep{estepp2014recovering}}
	\label{tab:glasses}
	\centering
	\scriptsize
	\setlength\tabcolsep{3pt} 
	\begin{tabular}{rc|ccccc}
	%\toprule
      \textbf{Method} &&  MAE & RMSE  & SNR & $\rho$ & WMAE  \\ \hline \hline
      Ours (LSTM) with Glasses && \textbf{2.17} & \textbf{4.55} & \textbf{7.33} & \textbf{0.87} & 0.21 \\
      CAN with Glasses && 3.33 & 6.56 & 3.80 & 0.76 & 0.24 \\
      Ours (LSTM) no Glasses && 2.55 & 5.79 & 4.68 & 0.59 & \textbf{0.20} \\
      CAN no Glasses && 2.57 & 5.13 & 2.50 & 0.66 & 0.22 \\
        %\bottomrule
  \end{tabular}
\end{wraptable}

\begin{figure*}[t!]
  \includegraphics[width=1\textwidth]{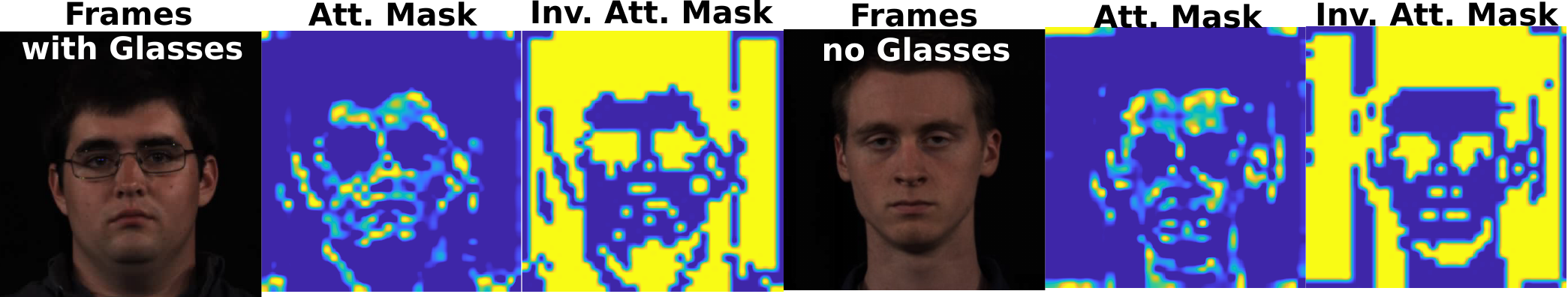}
  \caption{Comparison of attention masks and inverse attention masks on a video with and without glasses.} 
  \label{fig:glasses}
  \vspace{-0.5cm}
\end{figure*}

\textbf{Effect of Glasses.} We compared the performance of our denoising approach and the baseline CAN method on subjects with and without glasses. We found that our method offers largest improvements on subjects with glasses, as shown in Table~\ref{tab:glasses}. However, the attention masks output by CAN on subjects with and without glasses were comparable, as shown in Figure~\ref{fig:glasses}. Nine of the 25 subjects in the AFRL dataset were wearing glasses. No subjects in the MMSE-HR or MR-NIRP datasets were wearing glasses.

\end{document}

%% file: math_commands.tex
\usepackage{amsmath,amsfonts,bm}

\def\eqref#1{equation~\ref{#1}}
\def\1{\bm{1}}

\DeclareMathAlphabet{\mathsfit}{\encodingdefault}{\sfdefault}{m}{sl}
\SetMathAlphabet{\mathsfit}{bold}{\encodingdefault}{\sfdefault}{bx}{n}